\documentclass{article}
\usepackage{array}
\usepackage{amsmath}
\usepackage{latexsym}
\usepackage{amsfonts}
\usepackage{amssymb}
\usepackage[all]{xy}
\usepackage[latin5]{inputenc}
\usepackage{graphicx}
\usepackage{epsfig}
\def\d-{{$\Diamond$}\hspace{-.11in}{\small $^\mathbb{{\_\_}}$}}
\newcommand{\be}{\begin{equation}}
\newcommand{\ee}{\end{equation}}

\input epsf
\epsfverbosetrue

\begin{document}

\author{Oktay K. Pashaev and Oguz Yilmaz }
\title{Vortex Images and q-Elementary Functions}
\date{ Department of Mathematics,\\ Izmir Institute of Technology,\\ Izmir, 35430 Turkey}
\maketitle

\begin{abstract}
In the present paper problem of  vortex images in annular domain
between two coaxial cylinders is solved by q-elementary functions.
We show that all images  are determined completely as poles of
q-logarithmic function, where dimensionless parameter $q =
r^2_2/r^2_1$ is determined by square ratio
 of the cylinder radiuses.
The resulting solution for the complex potential is represented in
terms of the Jackson q-exponential function. By composing pairs of
q-exponents as the first Jacobi theta function and conformal
mapping to rectangular domain we link our solution with result of
Johnson and McDonald. We found that one vortex  cannot remain at
rest except at the geometric mean distance, but must orbit the
cylinders with constant angular velocity which is related with
q-geometric series. Vortices in two particular geometries in $q
\rightarrow \infty$ limit are studied.
\end{abstract}

\section{Introduction}
The classical method of images introduced by W. Thomson in 1845
becomes powerful method for solving boundary value problems in
electrostatics and hydrodynamics \cite{milnethomson},\cite{power}.
The method has been successfully applied to simple geometries as
spheres, cylinders and half-spaces, where explicit formulas have
been given. Unfortunately for complex body shapes the image
principle becomes extremely difficult even to find approximate
solution. This is why the image problem for which  solution can be
found  in an exact form, like merging cylinders \cite{palaniappan} for example,
becomes a member of very exclusive family . In the present paper we
consider exact solution of
 planar vortex problem in annular domain between two coaxial cylinders
 by method of images. This problem has many interesting
 applications. One of them is related to hydrodynamic interaction
 in which the modification of ambient flow by cylinders is
 carried out by obtaining the effect of single cylinder on the
 flow and then applying the boundary conditions on each cylinder
 to determine the unknown coefficients that appear in the series
 expansions. \cite{yilmaz} solved the diffraction problem of water
 waves by multiple cylinders placed at the free surface.

Another application is related with inviscid two-dimensional fluid dynamics experiments with
magnetized electron columns confined in a cylindrical trap\cite{Fajans}. The flow vorticity is proportional
to the electron density and the electric potential is analogous to the two dimensional stream
function. Thus, the electrons mimic ideal two dimensional fluid equations and by
creating electron columns with the appropriate density, one can model fluid flows.
It allows to model real problems of vortex interaction with topography \cite{johnson}.
Motion of a vortex in the neighborhood of a cosmic string \cite{gibbons} and influency on this
the cylindrically compactificated extra space dimensions is another class of possible
cosmological applications.

The problem of one vortex and one cylinder is connected with the
Circle Theorem of Milne-Thomson \cite{milnethomson} which can be
rewritten for the complex velocity of the flow $\bar V(z) = u_1 -
i u_2$ in the form \be \bar V (z) = \bar v(z) - \frac{r^2_1}{z^2}
\,v\left(\frac{r^2_1}{z}\right)\label{circle}\ee where $v(z)$ is
complex velocity of the flow in unbounded domain, and the second term in
(\ref{circle}) represents the correction to the complex velocity
by the cylinder of radius $r_1$ placed at the origin. For a vortex
at $z_0$, of strength $\kappa$ and circulation, $\Gamma = - 2\pi
\kappa$, (\ref{circle}) can be written explicitly as \be \bar V(z)
= \frac{i\kappa}{z-z_0} - \frac{i\kappa}{z-\frac{r^2_1}{\bar z_0}}
+ \frac{i\kappa}{z}\label{onecylinder}\ee where the second term represents
a vortex of strength $-\kappa$ at the inverse point of $z_0$, $\frac{r^2_1}{\bar z_0}$,
with respect to the cylinder. Henceforth, we shall call the
vortices at inverse points and at the centres of cylinders (or at
the infinity) "vortex images" or simply "images". Therefore, in
(\ref{onecylinder}), there are two images; one positive image at
the centre of the cylinder and another negative image at the
inverse point. In fact, images are used to replace the circle in
the infinite 2-D plane.

Another application is the case of a vortex at point $z_0$ inside a
cylindrical domain with radius $r_2$, C : $|z| < r_2$,

\be \bar V(z) = \frac{i\kappa}{z-z_0} -
\frac{i\kappa}{z-\frac{r^2_2}{\bar z_0}}\label{externaldisk}\ee
where the vortex image is located at point  $r^2_2/\bar z_0$ outside
C. The solution (\ref{externaldisk}) can be obtained from the circle
theorem by first using the mapping $z=1/\omega$ and also from the
Laurent series expansion of the solution.

The above two examples are limiting cases of the problem of a point
vortex in annular domain between two coaxial cylinders with inner
radius $r_1$ and outer radius $r_2$. We find in this case that the
solution is given as the infinite set of images in two cylinders,

\be \bar V(z)=   \sum_{n= - \infty}^{\infty} \left[
\frac{i\kappa}{z- z_0 q^n}- \frac{i\kappa}{z- \frac{r_1^2}{\bar
z_0}q^n}\right]. \ee

As we show in this paper these images are determined completely in
terms of q-logarithmic and q-exponential functions. Mathematical
study of these functions is connected with some applications in
the number theory, for calculation of the Euler's constant $\gamma$
by generalization of a classical formula due to Ramanujan and
Vacca \cite{SondowZudilin}, irrationality test \cite{Borwein2},
\cite{SondowZudilin}, the Stieltjes  transform of a positive
discrete measure \cite{Borwein1} and construction of the Pade
approximations \cite{Borwein1}. Another class of applications
related with physics is quantum groups and their representations
\cite{klimyk}. The physical systems with such symmetries started from
the quantum integrable systems and then extended to the several
q-deformed physical systems like the quantum linear harmonic
oscillator, generalized coherent states in quantum optics,
composite particle with the Chern-Simons flux - the anyons. It
was observed in general that physical systems with a fundamental
length scale have a symmetry of a quantum group \cite{celeghini1}.  In
nuclear physics the deformation parameter is related to the time
scale of strong interactions \cite{celeghini2}, while in solid state
physics with the lattice spacing \cite{bonechi}. Despite of this
particular progress, the direct interpretation of the deformation
parameter in these cases is sometimes incomplete or even
nonexistent. In the present paper we study the classical problem
of $N$ vortices placed in the annular region between two coaxial
cylinders with radii $r_1 < r_2$, and we see that the natural
dimensionless parameter $q = r^2_2/r^2_1$ plays the role of the
fundamental length
 and the infinite set of images produces the one dimensional q- lattice.

\section{$N$ Vortices in Annular Domain}
We consider the problem of $N$ point vortices in annular domain
$D:$ $\{r_1 \leq |z| \leq r_2\}$, where $z_1, ..., z_N$ are
positions of vortices with strengths $ \kappa_1,...,\kappa_N $
respectively. The region is bounded by two concentric circles:
$C_1: z \bar z = r_1^2$ and $ C_2: z\bar z= r_2^2 $

 The complex velocity is given by the Laurent series
\begin{equation}
\bar V(z)= \sum_{k=1}^{N} \frac{i \kappa_k}{z-z_k}+
\sum_{n=0}^{\infty} a_n z^n+ \sum_{n=0}^{\infty}
\frac{b_{n+1}}{z^{n+1}}\label{complex velocity}
\end{equation}
which has to satisfy the boundary conditions at both cylinders,
\be\left[\bar V(z)z + V(\bar z)\right]|_{C_k} = 0, \,
k=1,2.\label{bc}\ee The conditions (\ref{bc}) imply that no fluid
can penetrate any of the circular walls of the domain. If $\bf n$
denotes the normal to the boundary, the boundary condition is that
normal velocity must be zero, ${\bf u \cdot v} = u_n = 0$.

To find the unknown coefficients  we have to determine the boundary
conditions, \be\Gamma = \bar V(z)z +V(\bar z)\bar z =
\sum_{k=1}^{N}\frac{i \kappa_k z}{z-z_k}+ \sum_{n=0}^{\infty} a_n
z^{n+1}+ \sum_{n=0}^{\infty} \frac{b_{n+2}}{z^{n+1}} +
C.C.\label{eqn5.3}\ee where $C.C.$ stands for the complex conjugate.
Since on boundary $C_1$: $z\bar{z}={r_1}^2$ and $|z_k|>|z| $, we can
rewrite equation $(\ref{eqn5.3})$ as follows;
\begin{eqnarray}
\Gamma_{|_{C_1}}&=&\sum_{k=1}^{N} (-i \kappa_k)
\sum_{n=0}^{\infty}\left(\frac{z}{z_k}\right)^{n+1}+
\sum_{n=0}^{\infty} a_n z^{n+1}+ \sum_{n=0}^{\infty}
\frac{b_{n+2}}{z^{n+1}} + C.C.\nonumber \\
&=& \sum_{n=0}^{\infty} \left[ \sum_{k=1}^N \frac{-i
\kappa_k}{z_k^{n+1}}+a_n + \bar b_{n+2}\frac{1}{r_1^{2(n+1)}}
\right] z^{n+1} + C.C. =0.
\end{eqnarray}
 This implies the
following algebraic system
\begin{equation}
\sum_{k=1}^N \frac{-i \kappa_k}{z_k^{n+1}}+a_n + \bar
b_{n+2}\frac{1}{r_1^{2(n+1)}}=0, \,\,\,(n = 0,1,2...)
\quad\texttt{and}\quad b_1=0. \label{eqn1}\end{equation} Since on
boundary $C_2$: $z\bar{z}={r_2}^2$ and $|z_k|<|z| $, we can
rewrite equation (\ref{eqn5.3}) as follows;
\begin{eqnarray}
\Gamma_{|_{C_2}}&=& \sum_{k=1}^{N} (i \kappa_k)
\sum_{n=0}^{\infty}\left(\frac{z_k}{z}\right)^{n}+
\sum_{n=0}^{\infty} a_n z^{n+1}+ \sum_{n=0}^{\infty}
\frac{b_{n+2}}{z^{n+1}}+ C.C.\nonumber \\
&=& \sum^\infty_{n=0} \left[\sum_{k=1}^N  {\frac{-i \kappa_k \bar
z_k^{n+1}}{r_2^{2(n+1)}}}+a_n + \bar b_{n+2}\frac{1}{r_2^{2(n+1)}}
\right] z^{n+1} + C.C. =0.
\end{eqnarray}
This implies another algebraic system
\begin{equation}
\sum_{k=1}^N \frac{-i \kappa_k \bar z_k^{n+1}}{r_2^{2(n+1)}}+a_n +
\bar b_{n+2}\frac {1}{r_2^{2(n+1)}}=0, \,\,\,(n = 0,1 ,
2,...).\label{eqn2}\end{equation} We have two algebraic systems
$(\ref{eqn1})$ and $(\ref{eqn2})$. By substracting $(\ref{eqn2})$
from $(\ref{eqn1})$, we eliminate $a_n$
\begin{equation}
\bar b_{n+2}\left[\frac{1}{r_2^{2(n+1)}} -
\frac{1}{r_1^{2(n+1)}}\right]+ \sum_{k=1}^{N}(-i \kappa_k)
\left[\frac{\bar z_k^{n+1}}{r_2^{2(n+1)}} -
\frac{1}{z_k^{n+1}}\right]=0.
\end{equation}
If  $q \equiv r_2^2 / r_1^2$ is used we find
\begin{equation}
b_{n+2}= \sum_{k=1}^N \left( \frac{-i \kappa_k}{\bar z_k^{n+1}} \,
\frac{r_2^{2(n+1)}-|z_k|^{2(n+1)}}{q^{n+1}-1}\right)
\end{equation}
and from $(\ref{eqn1})$ we determine $a_n$,
\begin{equation}
a_n = \sum_{k=1}^N \frac{i \kappa_k}{z_k^{n+1}}- \bar
b_{n+2}\frac{1}{r_1^{2(n+1)}}, \end{equation} or
\begin{equation} a_n = \sum_{k=1}^N \frac{-i \kappa_k}{z_k^{n+1}}\, \frac{r_1^{2(n+1)}-
|z_k|^{2(n+1)}}{r_1^{2(n+1)}(q^{n+1}-1)}.
\end{equation}

The Taylor series part of $(\ref{complex velocity})$ gives the
following,
\begin{eqnarray}
\sum _{n=0}^{\infty} a_n z^n &=& \sum_{n=0}^{\infty} \sum_{k=1}^N
\frac{-i \kappa_k}{z_k^{n+1}}\, \frac{z^n}{q^{n+1}-1}-
\sum_{n=0}^{\infty} \sum_{k=1}^N \frac{(- i \kappa_k)
{\bar{z}}_k^{(n+1)}z^n}{r_1^{2(n+1)}(q^{n+1}-1)}\nonumber\\
&=& \sum_{k=1}^N \frac{-i\kappa_k}{z
(q-1)}\sum_{n=0}^{\infty}\frac{q -1}{q^{n+1} -
1}\left(\frac{z}{z_k}\right)^{n+1} +\sum_{k=1}^N
\frac{i\kappa_k}{z (q-1)}\sum_{n=0}^{\infty}\frac{q - 1}{q^{n+1} -
1}\left(\frac{z\bar z_k}{r^2_1}\right)^{n+1}
\nonumber \\
&=& \sum_{k=1}^N \frac{-i\kappa_k}{z (q
-1)}\sum_{n=0}^{\infty}\frac{1}{[n+1]}\left(\frac{z}{z_k}\right)^{n+1}
+\sum_{k=1}^N \frac{i\kappa_k}{z (q -
1)}\sum_{n=0}^{\infty}\frac{1}{[n+1]}\left(\frac{z\bar
z_k}{r^2_1}\right)^{n+1}
\nonumber \\
&=& \sum_{k=1}^N \frac{i \kappa_k}{z (q-1)} Ln_q \left(1-
\frac{z}{z_k}\right) -  \sum_{k=1}^N \frac{i \kappa_k}{z (q-1)} Ln_q
\left(1- \frac{z \bar z_k}{r_1^2}\right)\label{a_nz_n}
\end{eqnarray}
and the Laurent part gives,
\begin{eqnarray}
\sum _{n=0}^{\infty}\frac{b_{n+2}}{z^{n+2}}&=& \sum_{k=1}^N
\left(\frac{1}{z} \sum _{n=0}^{\infty} \frac{-i \kappa_k}{\bar
z_k^{n+1}} \frac{r_2^{2(n+1)}}{q^{n+1}-1}\frac{1}{z^{n+1}}\right) +
\sum_{k=1}^N \left(\frac{1}{z} \sum _{n=0}^{\infty}\frac{ i \kappa_k
z_k^{n+1}} {q^{n+1}-1}\frac{1}{z^{n+1}}\right)
\nonumber\\
&=& \sum_{k=1}^N \frac{-i\kappa_k}{z
(q-1)}\sum_{n=0}^{\infty}\frac{q-1}{q^{n+1} -
1}\left(\frac{r^2_2}{z\bar z_k}\right)^{n+1} +\sum_{k=1}^N
\frac{i\kappa_k}{z(q-1)}\sum_{n=0}^{\infty}\frac{q-1}{q^{n+1} -
1}\left(\frac{z_k}{z}\right)^{n+1}\nonumber \\
&=& \sum_{k=1}^N \frac{-i\kappa_k}{z
(q-1)}\sum_{n=0}^{\infty}\frac{1}{[n+1]}\left(\frac{r^2_2}{z\bar
z_k}\right)^{n+1} +\sum_{k=1}^N
\frac{i\kappa_k}{z(q-1)}\sum_{n=0}^{\infty}\frac{1}{[n+1] }\left(\frac{z_k}{z}\right)^{n+1}\nonumber \\
&=& \sum_{k=1}^N \frac{i \kappa_k}{z (q-1)} Ln_q \left(1-
\frac{r_2^2}{z \bar z_k}\right) -  \sum_{k=1}^N \frac{i \kappa_k}{z
(q-1)} Ln_q \left(1- \frac{z_k}{z}\right)\label{b_nz_n}
\end{eqnarray}
where $[n] = \frac{q^{n}-1}{q-1}$ and $Ln_q(1- x)\equiv -\sum_{n=
1}^{\infty}\frac{x^n}{[n]},\,\,|x| < q, \,\,q>1 \label{def2}$.

\section{Logarithmic and q-Exponential Functions}
By analogy with  ordinary logarithmic function : $|x| \leq 1$, $x
\neq -1$
\begin{equation}
\ln(1-x)=  -\sum_{n= 1}^{\infty}\frac{x^n}{n}
\end{equation}
q-logarithmic function is defined as

\begin{equation} Ln_q(1- x)\equiv -\sum_{n=
1}^{\infty}\frac{x^n}{[n]},\,\,|x| < q, \,\,q>1
\label{def2}.\end{equation} where q-number
\begin{equation}
[n] \equiv 1 + q + q^2 + ...+ q^{n-1} = \frac{q^{n}-1}{q-1}
\end{equation}
for any positive integer $n$.
 In the limiting case $q \rightarrow 1$
\begin{equation}
\lim_{q \rightarrow 1} [n] = n
\end{equation}

\begin{equation}
\lim_{q \rightarrow 1} Ln_q (1 - x) = \ln (1-x)
\end{equation}
Our definition relates with the q-Logarithmic function  of Borwein
\cite{Borwein1} \be L_q(x) = \sum_{n=1}^\infty \frac{x^n}{[n]}\ee
by \be Ln_q (1 - x) = -L_q(x)\ee and with that of Sondow and
Zudilin \cite{SondowZudilin}
\begin{equation}
\ln_q(1+x)= \sum_{n= 1}^{\infty}\frac{(-1)^{n-1}x^n}{q^n
-1},\,\,|x| < q, \,\,q>1 \label{def3}\end{equation} by
\begin{equation}
Ln_q (1 + x) = (q-1) \ln_q (1+ x)
\end{equation}
The q-derivative is defined as
\begin{equation}
D_q f(x) = \frac{f(qx) - f(x)}{(q-1)x}.\end{equation} Taking
q-derivative of $x^n$ and the q-logarithmic function, we get
\begin{equation}
D_q x^n = [n] x^{n-1}
\end{equation}
and
\begin{equation}
D_q Ln_q (1-x) = -\frac{1}{1- x} \label{lnderiv}
\end{equation}

For our problem, the following representation of q-Logarithmic
function \cite{Borwein1}, \cite{SondowZudilin} is crucial in the
complex domain: let $q$ be real, $q > 1$, then for $0 < |z| < q$
the following identity hold

\begin{equation}
Ln_q(1+z)=  \sum_{n= 1}^{\infty}\frac{(-1)^{n-1}z^n}{[n]}
=(q-1)\sum_{n= 1}^{\infty}\frac{z}{q^n+z}.\label{cor1}
\end{equation}
The proof is given by the following chain of transformations \be
\sum_{n= 1}^{\infty}\frac{z}{q^n+z} = z \sum_{n=
1}^{\infty}\frac{1}{q^n}\frac{1}{1+z\, q^{-n}} = z \sum_{n=
1}^{\infty}\frac{1}{q^n}
\sum_{k=1}^{\infty}\frac{(-z)^{k-1}}{q^{n(k-1)}} = \nonumber\ee
\be \sum_{k=1}^{\infty}(-1)^{k-1} z^k \sum_{n=1}^{\infty}
\frac{1}{q^{nk}}= \sum_{k=1}^{\infty}(-1)^{k-1} z^k
\frac{1}{q^{k}} \frac{1}{1 - q^{-k}} =
\sum_{k=1}^{\infty}\frac{(-1)^{k-1} z^k} {q^{k} -1} =\nonumber \ee
\be = \frac{1}{q-1}Ln_q (1+z)\ee

Next we will need Jackson's q-exponential functions defined as
\cite{Borwein1} \be E_q(z) = \sum_{n=0}^{\infty}
\frac{z^n}{[n]!}\label{qexp1}\ee \be E_q^{*}(z) =
\sum_{n=0}^{\infty} q^{n(n-1)/2}\frac{z^n}{[n]!}\label{qexp2}\ee
where $[n]! \equiv [1] [2] ... [n]$ and $q$ is real. As $q
\rightarrow 1$, both functions reduce to the ordinary exponential.
Application of q-derivative gives \be D_q E_q(x) =
E_q(x),\,\,\,\,D_q E_q^*(x) = E_q^*(q x)\label{derivexp}\ee The
function $E_q(z)$ is entire in $z$ if $|q| > 1$, while it has
radius of convergence $|1-q|^{-1}$ if $|q| < 1$. The function
$E_q^* (z)$ is entire for $|q| < 1$ and converges for $|z| < 1$ if
$|q| > 1$. These two functions are related by \be E_q(z) =
E_{1/q}^*(z)\label{expchangeq}\ee and \be E_q(-z) E_q^*(z) = 1\ee
Particularly important for us is the infinite product
representation \cite{Borwein1} \be E_q^*(z) = \prod_{k=0}^\infty
\left( 1 + z q^k (1-q)\right),\,\,\,\, |q| < 1\label{exprod1}\ee
and the related identity obtained by setting $q \rightarrow 1/q$
and shifting the argument \be \prod_{k=1}^\infty \left(1 -
\frac{z}{q^k}\right) = \frac{1}{1-z} E_q\left(\frac{-z}{1 -
q^{-1}}\right)\label{exprod}\ee which is entire for $|q| > 1$.

\subsection{Vortex Images and Poles of $q$-Logarithm}
Substituting equations $(\ref{a_nz_n})$ and $(\ref{b_nz_n})$ in
$(\ref{complex velocity})$ we get the following,
\begin{eqnarray}
\bar V(z)= \sum_{k=1}^N i \kappa_k \left[\frac{1}{z-z_k} +  \frac{1}{z(q-1)} \left[ Ln_q \left(1-\frac{z}{z_k}\right)- Ln_q \left(1- \frac{z\bar z_k}{r_1^2}  \right)+ Ln_q \left(1-\frac{r_2^2}{z \bar z_k}\right)-
Ln_q \left(1- \frac{z_k}{z} \right)\right]\right].\label{velocitylog}
\end{eqnarray}
It can be written in a more compact form in terms of q-derivatives with different basis q:
\be
\bar V(z)= \sum_{k=1}^N  \frac{i \kappa_k}{z-z_k} - \frac{i\kappa_k}{z_k}\left( [\alpha]_q D_{q^\alpha}Ln_q \left(1 - \frac{z}{z_k}\right)
- \frac{z_k^2}{z^2} [1-\alpha]_q D_{q^{1-\alpha}}Ln_q \left(1 - \frac{z_k}{z}\right)\right)
\ee
where the real q-number $[\alpha]_q = (q^{\alpha}-1)/(q-1)$, $\alpha = \alpha(k) \equiv \log_q (|z_k|^2/r^2_1)$.
Parameter $\alpha$ is restricted by $0 < \alpha < 1$. Particular values of this parameter correspond to different
positions of a vortex: 1) for $\alpha = 1$ the vortex is on the outer cylinder $|z_0| = r_2$,
2)  for $\alpha = 0$ the vortex is on the inner cylinder $|z_0| = r_1$ 3) for $\alpha = 1/2$ the vortex is at the geometric mean
distance  $|z_0| = \sqrt{r_1 r_2}$ (see Section 6 equation (\ref{geometricmean})) 4) for $\alpha = m/n$, where $m < n$ are positive
 integers, the vortex is at the generalized mean
distance  $|z_0|^n = r_1^{n-m} r_2^{m}$.

Expanding $q$-log according to $(\ref{cor1})$ we have
\begin{eqnarray}
\bar V(z)&=& \sum_{k=1}^N \frac{i \kappa_k}{z-z_k}+ \sum_{k=1}^N
\sum_{n=1}^{\infty} \frac{i \kappa_k}{z-z_kq^n}- \sum_{k=1}^N
\sum_{n=1}^{\infty} \frac{i \kappa_k}{z-q^n \frac{r_1^2}{\bar
z_k}}\nonumber
\\
&+& \sum_{k=1}^N \sum_{n=1}^{\infty} \left[\frac{i \kappa_k}{z}-
\frac{i \kappa_k}{z- q^{-n}\frac{r_2^2}{\bar z_k} }\right]-
\sum_{k=1}^N \sum_{n=1}^{\infty} \left[\frac{i \kappa_k}{z}-
\frac{i \kappa_k}{z- q^{-n} z_k }\right]
\end{eqnarray}
or
\begin{eqnarray}
\bar V(z)&=& \sum_{k=1}^N \frac{i \kappa_k}{z-z_k}+
\sum_{k=1}^N\left[ \sum_{n=1}^{\infty} \frac{i \kappa_k}{z-z_kq^n}+
\sum_{n=1}^{\infty} \frac{i \kappa_k}{z- z_k
q^{-n}}\right],\nonumber
\\
&-& \sum_{k=1}^N \left[\sum_{n=0}^{\infty} \frac{i \kappa_k}{z-
\frac{r_1^2}{\bar z_k}q^{-n}}  + \sum_{n=0}^{\infty} \frac{i
\kappa_k}{z- \frac{r_2^2}{\bar z_k}q^{n} } \right].\label{fcomplex}
\end{eqnarray}

Equation $(\ref{fcomplex})$ for complex velocity has countable
infinite number of pole singularities. These singularities can be
interpreted as vortex images in two cylindrical surfaces. For
simplicity let us consider only one vortex at position $z_0$, $r_1
< |z_0| < r_2$. Then the set of images in the cylinder $C_1$ we
denote $z_I^{(1)}, z_I^{(2)}, ...$ and in the cylinder $C_2$ as
$z_{II}^{(1)}, z_{II}^{(2)}, ...$, where
\begin{eqnarray}
z_I^{(1)}&=& \frac{r_1^2}{\bar z_0}
\,\,\,\,\,\,\,\,\,\,\,\,\,\,\,\,\,\,\,\,\,\,\,\,\,\,\,\,\,\,\,
z_{II}^{(1)}=
\frac{r_2^2}{\bar z_0} \\
z_I^{(2)}&=& \frac{r_1^2}{\bar
z_{II}^{(1)}}=\frac{z_0}{q}\,\,\,\,\,\,\,\,\,\,\,\,\, z_{II}^{(2)}=
\frac{r_2^2}{\bar z_{I}^{(1)}}=qz_0
\\
z_I^{(3)}&=& \frac{r_1^2}{\bar z_{II}^{(2)}}=\frac{r_1^2}{\bar
z_0}\frac{1}{q}\,\,\,\,\,\,\,\,\,\,\, z_{II}^{(3)}=
\frac{r_2^2}{\bar z_{I}^{(2)}}= \frac{r_2^2}{\bar z_0}q
\\
z_I^{(4)}&=& \frac{r_1^2}{\bar
z_{II}^{(3)}}=\frac{z_0}{q^2}\,\,\,\,\,\,\,\,\,\,\,\,\,\,
z_{II}^{(4)}= \frac{r_2^2}{\bar z_{I}^{(3)}}= z_0 q^2
\\
z_I^{(5)}&=& \frac{r_1^2}{\bar z_{II}^{(4)}}=\frac{r_1^2}{\bar
z_0}\frac{1}{q^2}\,\,\,\,\,\,\,\,\,\,\, z_{II}^{(5)}=
\frac{r_2^2}{\bar z_{I}^{(4)}}= \frac{r_2^2}{\bar z_0}q^2\\
&................&
\end{eqnarray}
Combining together and taking into account alternating signs (the
negative for the first image and the positive for the next one - the
image of the image ) we have two  sets of consecutive images
\begin{equation}
z_0,\,  z_I^{(1,-)},\, z_{II}^{(2,+)},\, z_I^{(3,-)},\,
z_{II}^{(4,+)},\, z_{I}^{(5,-)},...
\end{equation}
and
\begin{equation}
z_0,\, z_{II}^{(1,-)},\, z_I^{(2,+)},\,
z_{II}^{(3,-)},\, z_{I}^{(4,+)},\, z_{II}^{(5,-)},....
\end{equation}
This shows that the set of vortex images is completely determined
by simple pole singularities of the $q$-logarithmic function.
In the above representation (\ref{fcomplex}) by identity
$r^2_2/q^n = r^2_1/q^{n-1} $ we can combine sums so that, we have
\begin{eqnarray}
\bar V(z)&=& \sum_{k=1}^{N}\left[ \sum_{n= - \infty}^{\infty}
\frac{i \kappa_k}{z-z_k q^n}\right]- \sum_{k=1}^{N}\left[ \sum_{n=
0}^{\infty} \frac{i \kappa_k}{z- \frac{r_1^2}{\bar z_k}q^{-n}} +
\sum_{n=1}^{\infty}\frac{i \kappa_k}{z- \frac{r_1^2}{\bar
z_k}q^{n}}\right] \\
&=& \sum_{k=1}^{N} i \kappa_k \sum_{n= - \infty}^{\infty} \left[
\frac{1}{z- z_k q^n}- \frac{1}{z- \frac{r_1^2}{\bar z_k}q^n}\right].
\end{eqnarray}

\section{Complex Potential and q-Exponential function}

In this section we derive the complex potential of the flow
according to the relation

\be \bar V(z)= F'(z)\ee in terms of Jackson q-exponential function
(\ref{qexp1}). To construct $F(z)$ we use new function defined in
\cite{Borwein1} as \be f_q(z) \equiv \prod_{n=1}^{\infty} \left( 1
- \frac{z}{q^n}\right) = \frac{E_q\left( \frac{z q}{1-q}\right)}{1
- z} \label{fq}\ee where $|q| > 1$ and observe \be
\frac{f'_q(z)}{f_q(z)} = \frac{d}{dz}\ln f_q (z) = -
\sum_{n=1}^{\infty}\frac{q^{-n}}{1 - z q^{-n}}=
\sum_{n=1}^{\infty} \frac{1}{z -  q^{n}}.\label{flog}\ee Using
(\ref{cor1}) we have the relation with q-logarithmic function \be
\frac{Ln_q (1-z)}{(q-1)z} = \frac{d}{dz} \ln f_q(z) =
\frac{d}{dz}\ln \frac{E_q \left( \frac{q z}{1-q}\right)}{1-z}.\ee
This expression can be simplified if we use q-derivative of the
exponential function (\ref{derivexp}) by rescaling the argument
\be E_q \left( \frac{q z}{1-q}\right) = (1-z) E_q \left(
\frac{z}{1-q}\right)\label{expqderiv}\ee so that \be \frac{Ln_q
(1- \alpha z)}{(q-1)z} = \frac{d}{dz}\ln \,E_q \left( \frac{\alpha
z}{1-q}\right)\label{lnexp1}.\ee By similar arguments we find also
\be \frac{Ln_q (1- \frac{\alpha} {z})}{(q-1)z} = -\frac{d}{dz}\ln
\,E_q \left( \frac{\alpha}{(1-q) z}\right).\label{lnexp2}\ee If
now we apply these formulas to (\ref{velocitylog})  we get \be
\bar V(z) = \sum_{k=1}^N i \kappa_k \frac{d}{dz}\ln
\left[(z-z_k)\frac{E_q\left(\frac{z}{(1-q)z_k}\right)E_q\left(\frac{z_k}{(1-q)z}\right)}
{E_q\left(\frac{z \bar
z_k}{(1-q)r^2_1}\right)E_q\left(\frac{r^2_2}{(1-q)z \bar
z_k}\right)}\right]\label{vexp}.\ee Finally this implies complex
potential in the form \be F(z) = \sum_{k=1}^N i \kappa_k \left[\ln
(z - z_k) + \ln
\frac{E_q\left(\frac{z}{(1-q)z_k}\right)E_q\left(\frac{z_k}{(1-q)z}\right)}
{E_q\left(\frac{z \bar
z_k}{(1-q)r^2_1}\right)E_q\left(\frac{r^2_2}{(1-q)z \bar
z_k}\right)} \right]\label{fexp}.\ee The first term in the bracket
corresponds to the vortex at position $z_k$ while the second term
describes its images. All these images completely are determined
by zeroes of q-exponential functions. For the zeros of the
q-analogue of exponential function and asymptotic formulas for
varying parameter $q$ see \cite{qexpzero}.

\section{Conformal Mapping and Elliptic Functions}

To compare our solution (\ref{fexp}) with Johnson $\&$ Mc Donald
$(2004)$ we rewrite it in terms of elliptic functions. For
comparison purposes we fix the radius $r_2 = 1$ so that
 $q= \frac{r_2^2}{r_1^2}=\frac{1}{r_1^2}\equiv
\frac{1}{\tilde{q}^2}$, where we introduced new parameter $\tilde q
< 1$. Then according to (\ref{expchangeq}) $E_q(z) = E^*_{\tilde
q^2}(z)$ so that
 we find complex potential in terms of the second Jackson q-exponent
\be F(z) = \sum_{k=1}^N i \kappa_k \ln \left[
 (z-z_k) \frac{E^*_{\tilde q^2}(\frac{\tilde q^2 z }{(\tilde q^2 -1)z_k})
E^*_{\tilde q^2}(\frac{\tilde q^2 z_k }{(\tilde q^2
-1)z})}{E^*_{\tilde q^2}(\frac{z \bar z_k }{(\tilde q^2 -1)})
E^*_{\tilde q^2}(\frac{ \tilde q^2}{(\tilde q^2 -1)z \bar
z_k})}\right].\label{phiexp2} \ee Using representation of
q-exponentials as an infinite products (\ref{exprod1}) we have \be
E^*_{\tilde q^2}\left(\frac{z }{\tilde q^2 -1}\right) =
\prod^\infty_{n=0} \left(1 - \tilde q^{2n} z\right) \equiv
\left(1-z\right)^\infty_{\tilde q^2} \label{prod1}\ee \be
E^*_{\tilde q^2}\left(\frac{\tilde q^2 z }{\tilde q^2 -1}\right) =
\prod^\infty_{n=1} \left(1 - \tilde q^{2n} z\right) =
\frac{\left(1-z\right)^\infty_{\tilde q^2}}{1-z}. \label{prod2}\ee
The first Jacobi theta function is defined as an infinite product
\cite{WW} \be \Theta_1 (x; \tilde{q}) = 2\,G
\,\tilde{q}^{\frac{1}{4}}\, \sin x
\,\prod_{n=1}^{\infty}\left(1-\tilde{q}^{2n}e^{2ix}\right)
\left(1-\tilde{q}^{2n}e^{-2ix}\right)\ee where \be G \equiv
\prod_{n=1}^{\infty}(1-\tilde{q}^{2n}) \ee
($\tilde q
< 1$) or

\be \Theta_1 (x; \tilde{q}) = \frac{G
\,\tilde{q}^{\frac{1}{4}}}{2\,\sin x}
\,\left(1-e^{2ix}\right)^\infty_{\tilde q^2}
\left(1-e^{-2ix}\right)^\infty_{\tilde q^2}.\label{teta1}\ee

This theta function is composed from two q-exponentials

\be \Theta_1 (x; \tilde{q}) = 2\,G \,\tilde{q}^{\frac{1}{4}}\,
\sin x \, E^*_{\tilde q^2}\left(\frac{\tilde q^2 \,e^{2i x}
}{\tilde q^2 -1}\right) E^*_{\tilde q^2}\left(\frac{\tilde q^2
\,e^{-2i x} }{\tilde q^2 -1}\right).\label{teta2}\ee

Then complex potential becomes

\be F(z) = \sum_{k=1}^N i \kappa_k \ln \left[
 z \frac{\left(1 - \frac{1}{z \bar z_k}\right)}{\left(1 - \frac{z}{z_k}\right)}
 \frac{\left(1- \frac{z}{z_k}\right)^\infty_{\tilde q^2} \, \left(1- \frac{z_k}{z}\right)^\infty_{\tilde q^2}}
 {\left(1- z \bar z_k \right)^\infty_{\tilde q^2} \, \left(1- \frac{1}{ z \bar z_k}\right)^\infty_{\tilde q^2} }\right].\label{fproduct}
\ee

If we denote \begin{equation} \frac{z}{z_k}= e^{2iu_k} ,\,\,\,\,\,
z \bar z_k=e^{2iv_k},\,\,\,\,\, z_k \bar z_k=
\frac{e^{2iv_k}}{e^{2iu_k} }\end{equation} where $k = 1,\cdots,N$,
we obtain,
\begin{equation}
F(z)= \sum_{k=1}^N i \kappa_k \ln \left[  z_k\, e^{2i
u_k}\frac{\left(1 - e^{-2i v_k}\right)}{\left(1 - e^{2i u_k}\right)}
 \frac{\left(1- e^{2i u_k}\right)^\infty_{\tilde q^2} \, \left(1- e^{-2i u_k}\right)^\infty_{\tilde q^2}}
 {\left(1- e^{2i v_k} \right)^\infty_{\tilde q^2} \, \left(1- e^{- 2i v_k}\right)^\infty_{\tilde q^2}
 }\right].
\end{equation}
By using (\ref{teta1}) we have

\begin{equation}
F(z)= \sum_{k=1}^N i \kappa_k  \left[\ln\left[ \frac{\Theta_1(u_k,
\tilde q)}{\Theta_1 (v_k, \tilde q)}\right] + \ln
\left[-\left(\frac{z_k}{\bar z_k}\right)^{1/2}\right]\right].
\end{equation}
In terms of coordinates $\tau \equiv -\ln z$, $\tau_k \equiv -\ln
z_k$  and $ \bar \tau_k \equiv -\ln \bar z_k$, conformally mapping the annulus in the $z$ plane to a
rectangle in the $\tau$ plane,  finally we find
\begin{equation}
F(z)= \sum_{k=1}^N i \kappa_k \ln \left[\frac{\Theta_1 (i\frac{\tau-
\tau_k}{2}, \tilde{q})}{\Theta_1 (i\frac{\tau+ \bar\tau_k}{2},
\tilde{q})}\right] + F_0
\end{equation}
where $F_0$ is a real constant and branch for logarithm is chosen
such that $\ln(-1)=i\pi$, \be F_0 = -\frac{i}{2}\sum_{k=1}^N
\kappa_k (\tau_k - \bar \tau_k) - \pi \sum_{k=1}^N \kappa_k.\ee

For the stream function, we have
\begin{equation}
\Psi = \frac{F-\bar F}{2i}= \sum_{k=1}^N\kappa_k \ln \left|
\frac{\Theta_1 (\frac{i}{2}(\tau- \tau_k), \tilde{q})}{\Theta_1
(\frac{i}{2}(\tau+ \bar\tau_k), \tilde{q})}\right|.
\end{equation}
This coincides with the result of Johnson $\&$ Mc Donald $(2004)$.
\section{Motion of a Point Vortex in Annular Domain}

We use the above formulas to determine the motion of a single
vortex in annular domain. Complex velocity at the vortex position
is determined by, \be \dot{z}_0 = \dot{x}_0 + i \dot{y}_0 =
V_0(\bar z)|_{z=z_0} \ee where in the complex velocity
\begin{eqnarray}\bar V_0(z)& =& \frac{i\kappa}{z (q-1)}\left[ Ln_q \left(1 - \frac{z}{z_0}  \right) - Ln_q \left(1 - \frac{z_0}{z}  \right)
+ Ln_q \left(1 - \frac{r^2_2}{z\bar z_0}  \right) - Ln_q \left(1 - \frac{z \bar z_0}{r^2_1}  \right)\right]\nonumber\\
& = &\sum_{n= \pm 1}^{\pm\infty}\frac{i \kappa}{z-z_0 q^n}- \sum_{n=
\pm 1}^{\pm\infty}\frac{i \kappa}{z- \frac{r_1^2}{\bar z_0}
q^n}.\nonumber
\end{eqnarray}
contribution of the vortex itself is excluded. If we take into account that q-harmonic series \cite{qharmonic}
\be H(q) \equiv \sum_{n=1}^\infty \frac{1}{[n]} = - Ln_q 0\ee
converges for $q > 1$ then at $z = z_0$ the first two terms cancel each other and we get the following equation of motion \be \dot z_0 =
\frac{i\kappa}{\bar z_0 (q-1)}\left[Ln_q \left(1 -
\frac{|z_0|^2}{r_1^2}  \right) - Ln_q \left(1 -
\frac{r_2^2}{|z_0|^2}  \right)\right].\ee The last equation gives,
\begin{equation}
\bar z_0 \dot{z}_0 + z_0 \dot{\bar z}_0=\frac{d}{dt}|z_0|^2=0
\rightarrow |z_0|= \mathrm{const}.
\end{equation}
This implies that the distance of the vortex from the origin is a
constant of motion. Then only the argument of $
z_0=|z_0|e^{i \varphi(t)}$, is a time dependent
function,
\begin{equation}
\varphi(t)= \omega t + \varphi_0
\end{equation}
where constant frequency $\omega$ dependens on modulus $|z_0|$,
\begin{eqnarray}
\omega &=& \frac{\kappa}{|z_0|^2 (q-1)}\left[ Ln_q \left( 1-
\frac{|z_0|^2}{r_1^2}\right)-  Ln_q \left( 1-
\frac{r_2^2}{|z_0|^2}\right) \right].\label{freqlog}
\end{eqnarray}
So we find that the vortex uniformly rotates around the origin,
\begin{equation}
z_0(t)= |z_0| e^{i \omega t + i\varphi_0}= z_0(0) e^{i \omega t}.
\end{equation}
with frequency depending on the vortex strength, the initial position and
geometry of the annular domain. This reflects
the fact that the motion of vortex results from interaction with an
infinite set of its images in the cylinders. The frequency
(\ref{freqlog}) vanishes when
\begin{equation}
 Ln_q \left( 1-
\frac{|z_0|^2}{r_1^2}\right)= Ln_q \left( 1-
\frac{r_2^2}{|z_0|^2}\right)
\end{equation}
or $|z_0|^4 = r_1^2 r_2^2$. It means that at the geometric mean
distance
\begin{equation}
|z_0| = \sqrt{r_1 r_2}
\label{geometricmean}\end{equation}
vortex is at the rest. This equation has simple geometrical meaning that any two intersection points
of cylinders with a ray (say $r_1 e^{i\alpha}$ and $r_2 e^{i\alpha}$) are images of each other in a cylinder with radius
 $\sqrt{r_1 r_2}$.
At this distance angular velocity changes the
sign and when the vortex approaches the cylinders, $\omega$
grows in modulus; $\omega \rightarrow \kappa/(2r^2_1 \epsilon)$
when $|z_0| \rightarrow r_1 e^\epsilon \approx r_1 (1 +
\epsilon)$, $0< \epsilon << 1$, and $\omega \rightarrow
-\kappa/(2r^2_2 \epsilon)$ when $|z_0| \rightarrow r_2
e^{-\epsilon} \approx r_2 (1 - \epsilon)$, $0< \epsilon << 1$.

Here we like to indicate an intriguing relation of our solution
 with number theory. The frequency (\ref{freqlog}) is combination
 of two opposite sign frequencies, $\omega = \omega_1 + \omega_2$,
 made from the q-logarithm functions
\be \omega_1= \frac{\kappa}{|z_0|^2 (q-1)} Ln_q \left( 1-
\frac{|z_0|^2}{r_1^2}\right) = \frac{-\kappa}{|z_0|^2 (q-1)}
\sum_{n=1}^{\infty}
\frac{\left(\frac{|z_0|^2}{r_1^2}\right)^n}{[n]} \label{freq1}\ee
\be \omega_2=- \frac{\kappa}{|z_0|^2 (q-1)} Ln_q \left( 1-
\frac{r_2^2}{|z_0|^2}\right)= \frac{\kappa}{|z_0|^2 (q-1)}
\sum_{n=1}^{\infty}
\frac{\left(\frac{|r_2^2}{|z_0|^2}\right)^n}{[n]} \label{freq2}\ee
The last representation shows that every frequency is infinite
superposition of frequencies coming from every vortex image.
Moreover for $|z_0| = r_1$ the frequency $\omega_1 = H(q)$, while
for $|z_0| = r_2$ the frequency $\omega_2 = -H(q)$ (for simplicity
we took coefficients equal one), where $H(q)$ is the q-harmonic
series. Contribution of N images in the frequency at these
limiting cases are given by q-harmonic numbers
\cite{qharmonicnumber} \be \omega^{(N)}= H_N(q) = \sum_{n=1}^N
\frac{1}{[n]_q} \ee The frequencies (\ref{freq1}) and
(\ref{freq2}) are compensating each other at the geometric mean
distance (\ref{geometricmean}). In the annular region $r_1 < |z_0|
< \sqrt{r_1 r_2} $, $\omega_1 > |\omega_2|$ and resulting $\omega
> 0$, while in the region $\sqrt{r_1 r_2} < |z_0| < r_2
 $, $\omega_1 < |\omega_2|$ and resulting $\omega
< 0$. If we consider geometry with parameter $q \geq 2 $ and the
unit strength vortex at the distance such that all arguments are
non-zero rational, then problem of rationality of the frequency of
motion is related with problem of rationality of q-logarithms.
Starting from early result of Erdos it was proved that the last
one is irrational \cite{Borwein2}. We can expect influence of this
irrationality on the character of multiple vortex dynamics.

\section{The One Vortex Problem in $q \rightarrow \infty$ Limiting Cases}

Usually in applications of q-calculus the limit $q \rightarrow 1$
corresponds to  reduction of the q-elementary functions to the
standard elementary functions. However in our problem this limit
corresponds to $r_1 = r_2$ and the region reduces to the circle.
More interesting is the limit when $q \rightarrow \infty$.
To study this limit we need corresponding limits of q-elementary
functions. When the q-logarithm is expanded we get the following \be
Ln_q (1 + z) = (q-1) \sum^\infty_{n=1} \frac{z}{q^n +z} = (1 -
\frac{1}{q})\sum^\infty_{n=1}\frac{z}{q^{n-1}}(1 - \frac{z}{q^n} +
...).\ee This implies that \be \lim_{q \rightarrow \infty} Ln_q (1
+ z) = z.\label{lninfty} \ee Using the $q-$ derivative of $Ln_q(1 - z)$
function (\ref{lnderiv}) \be \frac{Ln_q(1 - q z) - Ln_q(1 -
z)}{(q-1)z} = D_q Ln_q (1 - z) = - \frac{1}{1 - z}\ee and
(\ref{lninfty}) we get another limit \be \lim_{q \rightarrow
\infty}\frac{Ln_q(1 - q z)}{q - 1} = \frac{z}{z -
1}.\label{lnqinfty}\ee For the large $q >> 1$ we have \be [n] = \frac{q^n -
1}{q - 1} \approx q^{n-1}\ee \be [n]! = [1]\cdot[2]\cdot[3]...[n]
\approx q\cdot q^2\cdot q^3 ... q^{n-1} = q^{n(n-1)/2}\ee and \be
E_q(z) = \sum^\infty_{n=0} \frac{z^n}{[n]!} \approx
\sum^\infty_{n=0} \frac{z^n}{q^{n(n-1)/2}}= 1 + z + \frac{z^2}{q}
+ ...,\ee so that \be \lim_{q \rightarrow \infty} E_q (z) = 1 +
z.\label{expinfty}\ee Applying the q-derivative \be \frac{E_q(q z) -
E_q(z)}{(q-1)z} = D_q E_q (z) = E_q (z),\ee we have identities \be
E_q\left(\frac{q z}{q-1}\right) = (1+z)E_q
\left(\frac{z}{q-1}\right)\label{eq1}\ee \be
E_q\left(\frac{z}{q}\right) = \frac{E_q (z)}{1 + \frac{q-1}{q}
z}.\label{eq2}\ee In the limit $q \rightarrow \infty $ they imply
that \be \lim_{q \rightarrow \infty}E_q\left(\frac{z}{q}\right) =
1 \label{limexp1}\ee \be \lim_{q \rightarrow
\infty}E_q\left(\frac{q z}{q-1}\right) = 1 + z. \label{limexp2}\ee
Since $q = r^2_2/r^2_1$, for the limit $q \rightarrow \infty$ we have two different geometrical cases:

1) when $r_1 = constant$ and  $r^2_2 = q r^2_1 \rightarrow \infty$,
the outer cylinder grows unlimited so that we have just the one
cylinder problem. Complex velocity (\ref{velocitylog}) by replacement $r^2_2 = q
r^2_1$, taking the limit $q \rightarrow \infty$ and using
(\ref{lninfty}),(\ref{lnqinfty}) gives finite result \be \bar V(z)
= \frac{i\kappa}{z-z_0} - \frac{i\kappa}{z}\frac{r^2_1/\bar z_0}{z
- r^2_1/\bar z_0}\ee which is exactly the result of the circle
theorem (\ref{onecylinder}). For angular velocity we have using
(\ref{freqlog}) and the limits
(\ref{lninfty}),(\ref{lnqinfty}),\be \omega = \kappa
\frac{r^2_1}{|z_0|^2(|z_0|^2 - r^2_1)}.\ee For the complex
potential we have \be F(z) =  i \kappa \ln \left[(z - z_0)
\frac{E_q\left(\frac{z}{(1-q)z_0}\right)E_q\left(\frac{z_0}{(1-q)z}\right)}
{E_q\left(\frac{z \bar z_0}{(1-q)r^2_1}\right)E_q\left(\frac{q
\,r^2_1}{(1-q)z \bar z_0}\right)} \right].\label{fexpinfty1}\ee
Applying the limits (\ref{limexp1}),(\ref{limexp2}) we obtain two
images as required by the circle theorem \be F(z) =  i \kappa
\left[\ln (z - z_0) - \ln \left(z - \frac{r^2_1}{\bar z_0} \right)
+ \ln z \right].\label{flim1}\ee

2) when $r_2 = constant$ and $r^2_1 = r^2_2/q \rightarrow 0$, the
inner cylinder decreases until the thin string and then disappears.
In this limiting case in complex velocity (\ref{velocitylog}) we
replace $r^2_1 = r^2_2/q $, and use formulas
(\ref{lninfty}),(\ref{lnqinfty}) so that we get expression
(\ref{externaldisk}). For angular velocity of one vortex from
(\ref{freqlog}) by limits (\ref{lninfty}),(\ref{lnqinfty}) we have
\be \omega = -\kappa \frac{1}{r^2_2 -|z_0|^2 }.\ee Applying above limits
(\ref{limexp1}),(\ref{limexp2}) to the complex
potential  \be F(z) =  i \kappa \ln \left[(z - z_0)
\frac{E_q\left(\frac{z}{(1-q)z_0}\right)E_q\left(\frac{z_0}{(1-q)z}\right)}
{E_q\left(\frac{q \,z \bar
z_0}{(1-q)r^2_2}\right)E_q\left(\frac{\,r^2_2}{(1-q)z \bar
z_0}\right)} \right].\label{fexpinfty2}\ee  we obtain just one image as expected
from (\ref{externaldisk}) \be F(z) =  i \kappa \left[\ln (z - z_0)
- \ln \left(z - \frac{r^2_2}{\bar z_0} \right) + \ln
\left(-\frac{r^2_2}{\bar z_0} \right) \right].\label{flim1}\ee

\section{Conclusions}

We have shown that the infinite set of images for a vortex in
annular domain between two coaxial cylinders is given completely in
terms of q-logarithmic function for complex velocity and in terms of the
Jackson q-exponential function for the complex potential. Recent
results on Pade approximation for the q-elementary functions
\cite{Borwein1} could be efficient approximation in the vortex image
problem, restricting number of images. For example in paper
\cite{SondowZudilin} in order to compute the q-logarithm, with
reference on P. Sebah, the next formula proposed \be ln_q (1+z) = z
\sum_{n=1}^N \frac{1}{q^n + z} + r_N(z)\ee where \be r_N(z) =
\sum_{n=1}^\infty \frac{(-1)^{n-1}z^n}{q^{Nn (q^n -1)}}\ee
with $N$ being any positive integer.
Then application of this formula to our problem gives N-vortex images
approximation which could be useful in applications.

By conformal mapping of the annular region to the rectangular one and composing q-exponents, the stream function has been
represented in terms of the Jacobi elliptic function and it coincides
exactly with result of Johnson, Mc Donald. Since annular region
can be conformally mapped to exterior of two cylinders in the
plane, our solution provides also solution of the last problem
in terms of q-elementary functions. Results of calculations are
in preparation \cite{Cagatay}.

Finally we like to note that
elementary relation between the hydrodynamical problem and
q-calculus considered in this paper could be applied to several
physical situations with the same geometry, as the electrostatic
problem or the problem of anyons \cite{pashaev}. The image
picture allow us to construct Green function in the domain in
terms of q-elementary functions and apply it to other problems like
the vortex dynamics.

\section{Acknowledgements}
This work was partially supported by TUBITAK under the Grant No.
106T447 and by Izmir Institute of Technology.

\end{document}